# The growth equation of cities

Vincent Verbavatz[1,2], Marc Barthelemy[1,3,*]

**Affiliations:**

[1]Université Paris-Saclay, CNRS, CEA, Institut de physique théorique, 91191, Gif-sur-Yvette, France.

[2]École des Ponts ParisTech, Champs-sur-Marne, France.

[3]Centre d'Etude et de Mathématique Sociales, CNRS/EHESS, 54 Boulevard Raspail, 75006 Paris, France.

*Correspondence to: marc.barthelemy@ipht.fr

**Summary paragraph**

The science of cities seeks to understand and explain regularities observed in the world's major urban systems. Modelling the population evolution of cities is at the core of this science and of all urban studies. Quantitatively, the most fundamental problem is to understand the hierarchical organization of cities and the statistical occurrence of megacities, first thought to be described by a universal law due to Zipf [1, 2], but whose validity has been challenged by recent empirical studies [3, 4]. A theoretical model must also be able to explain the relatively frequent rises and falls of cities and civilizations [5], and despite many attempts [6, 7, 8, 9, 10] these fundamental questions have not been satisfactorily answered yet. Here we fill this gap by introducing a new kind of stochastic equation for modelling population growth in cities, which we construct from an empirical analysis of recent datasets (for Canada, France, UK and USA) that reveals how rare but large interurban migratory shocks dominate city growth. This equation predicts a complex shape for the city distribution and shows that Zipf's law does not hold in general due to finite-time effects, implying a more complex organization of cities. It also predicts the existence of multiple temporal variations in the city hierarchy, in agreement with observations [5]. Our result underlines the importance of rare events in the evolution of complex systems [11] and at a more practical level in urban planning.

**Main text**

**Growth of Cities and Zipf's law**

Constructing a science of cities has become a crucial task for our societies that grow always more concentrated in urban systems. A better planning could be achieved with such a science that seeks to understand city growth and how it affects society and environment [12]. Various important aspects of cities such as urban sprawl, infrastructure development or transport planning depend on the population evolution over time, and multiple theoretical attempts were made in order to understand this crucial phenomenon. To this day, most of these works were done with the idea that the stationary state for a set of cities is described by Zipf's law. This law is considered as a cornerstone of urban economics and geography [3], and states that the population distribution of urban areas on a given territory (or country) displays a Pareto law with exponent equal to 2 or, equivalently, that the city populations sorted in decreasing order versus their ranks follow a power-law with exponent 1. This allegedly regularity through time and space is probably the most striking fact in the science of cities and has triggered intense debates and many studies for more than a century [1, 2, 5, 10, 13, 14, 15, 16, 17, 18] [19, 20, 21, 22, 23, 24, 25, 26, 27, 28]. Indeed, this result characterizes the hierarchical organization of cities, and in particular quantifies the statistical occurrence of large cities. Zipf's law implies that in any country the city with the largest population is generally twice as large as the next-biggest, and so on. It is a signature of the very large heterogeneity of city sizes and shows that cities are not governed by optimal considerations that would lead to one unique size but, on the contrary, that cities sizes are broadly distributed and follow some sort of hierarchy [16]. The empirical value of the Pareto exponent informs us about the hierarchical degree of a system of cities: a large value of the exponent corresponds to a more equally distributed population among cities and on the contrary, for small exponent values, the corresponding system of cities is very heterogeneous with a few megacities.

Studies in economics suggested that Zipf's law is the result of economic shocks and random growth processes [6, 7, 8]. Gabaix [10] proved in a seminal paper how Gibrat's law of random growth [9] - that assumes a population growth rate independent from the size of the city - can lead to a Zipf law with exponent 1, at the expense of the additional and untested assumption that cities cannot become too small. This model remains so far the most accepted paradigm to understand city growth. Since then, it has also been understood on simplified theoretical models (without any empirical arguments) that migrations from other cities or countries are determinant in explaining random growth [29]. However, while most of these theoretical approaches focused on how to explain Zipf's law with exponent 1, recent empirical studies [3, 4] boosted by the increased number of data sources have questioned the existence of such a universal power-law and have shown that Zipf's exponent can vary around 1 depending on the country, the time period, the definition of cities used, or the fitting method [13, 21, 30, 31] (we illustrate this in the Extended data Figure 1 showing that no universal result for the population distribution is observed) leading to the idea that there is no reason to think that Zipf's law holds in all cases [32].

Beyond understanding the stationary distribution of urban populations, lies also the problem of their temporal evolution. As already noted in [5], the huge number of studies about the population distribution contrasts with the few analysis of the time evolution of cities. As discussed in [5], cities and civilizations are rising and falling many times on a large range of time scales, and Gabaix's model is both quantitatively and qualitatively unable to explain these specific chaotic dynamics.

So far, a model able to explain simultaneously observations about the stationary population distribution and the temporal dynamics of systems of cities is therefore missing. In particular, we are not able at this point to identify the causes of the diversity of empirical observations about the hierarchical organization of cities, the occurrence of megacities, and the empirical instability in city dynamics with falls or births of large cities on short time scales. In this respect, we do not need just a quantitative improvement of models but a shift of paradigm. In this paper, we show that city growth is dominated by rare events, namely large interurban migratory shocks, rather than by the average growth rate. Rare but significant positive or negative migratory flows can destabilize the hierarchy and the dynamics of cities on very short time scales, leading to the chaotic dynamics of cities observed through History. Based on the empirical analysis of migrations flows in four countries, we derive in the following the new stochastic equation of city growth able to explain empirical observations about both their statistics and their temporal dynamics.

**Deriving the equation of city growth**

To understand city growth, we need a robust bottom-up approach, starting from elementary mechanisms governing the evolution of cities. Without loss of generality, the growth dynamics of a system (such as a country) of cities $i$ of size $S_i$ can be decomposed into the sum of an inter-urban migration term between metropolitan areas and an 'out-of-system' term that combines the other sources of growth: natural growth (birth and deaths) and migrations that do not occur within the system of cities (international migrations and exchanges with smaller towns). We denote by $N(i)$ the set of neighbours of city $i$, ie. cities that exchange a non-zero number of inhabitants. On the four recent datasets of migrations that we use here (USA for 2012-2017, France for 2003-2008, UK for 2012-2016 and Canada for 2012-2016) we find for France and USA that $N(i) \vee S_i^\gamma$ where $\gamma \approx 0.5$ (Extended data Figure 2). The British and Canadian datasets are fully-connected leading to $\gamma = 0$. The time evolution of the population size $S_i$ can then be written as:

$$\partial_t S_i = \eta_i S_i + \sum_{j \in N(i)} J_{j \to i} - J_{i \to j} \qquad (1)$$

where the quantity $\eta_i$ is an uncorrelated random variable accounting for the 'out-of-system' growth of city $i$ and that we found in the data to be gaussian distributed (Extended Data Figure 3). The flow $J_{i \to j}$ is the number of individuals moving from $i$ to $j$ during the time $dt$. If there is an exact balance of migration flows ($J_{i \to j} = J_{j \to i}$), the equation becomes equivalent to Gibrat's model [9] which predicts a lognormal distribution of populations.

Starting from this general equation (1) is very natural as it amounts to write the balance of births, deaths and migrations but - as it is often the case of general basic equations - it is very difficult to use for making predictions. Simplifications of this equation were proposed in [29] where various assumptions (such as the gravity model for migration for example) lead to Gibrat's model but however missed the very large fluctuations of migrations, a crucial ingredient as we will see below. We also note that this general stochastic equation (1) was discussed in another context [33] and is a central object in the statistical physics of disordered systems. Regarding cities, the migration flow $J_{i \to j}$ depends a priori (and at least) on the populations $S_i$ and $S_j$ and the distance $d_{ij}$ between cities $i$ and $j$. Using a standard gravitational model [34, 35], we show that for France and USA, the dominant contribution to the flow $J_{i \to j}$ comes from the populations and that the role of distance appears as a second-order effect (see SI for details). This result suggests that the $J_{i \to j}$ term can be represented by a variable of the form $I_0 S_i^\mu S_j^\nu x_{ij}$ where the random variables $x_{ij}$ have an average equal to 1 and encode all the noise and multiple effects, including distance. We denote by $I_{ji} = J_{i \to j}/S_i$ the probability per unit time and per capita to move from city $i$ to city $j$. The left panel of Fig. 1 shows that the ratio $I_{ij}/I_{ji}$ versus the ratio of populations $S_i/S_j$ displays a linear behavior on average. This implies that $\mu = \nu$, and that we have on average a sort of detailed balance $J_{i \to j} \geq J_{j \to i} >$, but that crucially, fluctuations are non-zero. More precisely, if we denote by $X_{ij} = \frac{J_{i \to j} - J_{i \to j}}{I_0 S_i^\nu}$, we observe that these random variables $X_{ij}$ are heavy-tailed, i.e. they are distributed according to a broad-law that decreases asymptotically as a power-law with exponent $\alpha < 2$ (see SI for more details and empirical evidence). The sum in the second term of the r.h.s. of Eq. (1) can then be rewritten as

$$\sum_{j \in N(i)} J_{j \to i} - J_{i \to j} = I_0 S_i^\nu \sum_{j \in N(i)} X_{ij} \qquad (2)$$

and according to the generalized version of the central limit theorem [36] (assuming that correlations between the variables $X_{ij}$ are negligible), the random variable

$$\zeta_i = \frac{1}{N(i) \vee^{1/\alpha}} \sum_{j \in N(i)} X_{ij}$$

follows (for a large enough $N(i)$) a Lévy stable law $L_\alpha$ of parameter $\alpha$ and scale parameter $s$. This is empirically confirmed in Figure 1 (right panel): French, American, British and Canadian data are better fitted by a Lévy stable law than by any other distribution and the estimates of $\alpha$ (using different methods) are given in Table 1. We are thus led to the conclusion that the growth of systems of cities is governed by a stochastic differential equation (SDE) of a new type with two independent noises:

$$\partial_t S_i = \eta_i S_i + D S_i^\beta \zeta_i \qquad (3)$$

where $D = sI_0$, $\beta = \nu + \frac{\gamma}{\alpha}$ and where $\eta_i$ is a gaussian noise of mean the average growth rate $r$ and dispersion $\sigma$. This is the growth equation of cities that governs the dynamics of large urban populations and which is our main result here. In this equation both noises are uncorrelated, multiplicative and Itô's convention [37] seems here to be the more appropriate as population sizes at time $t$ are computed independently from inter-urban migrations terms at time $t + dt$. Estimates for the various parameters together with the prediction for the value of $\beta$ are given in Table 2.

The central limit theorem together with the broadness of inter-urban migration flow allowed us to show that many details in Eq. (1) are in fact irrelevant and that the dynamics can be described by the more universal Eq. (3). Starting from the exact equation (1) is therefore not only doomed to failure in general, but is also not useful. The importance of migrations was already sensed in [29], but the authors derived a stochastic differential equation with multiplicative Gaussian noise, which we show here to be incorrect: we have indeed a first multiplicative noise term but also crucially another term that is a multiplicative Lévy noise with zero average. This is a major and non-trivial theoretical shift that was missed in all previous studies on urban growth and which has many capital implications, both in understanding stationary and dynamic properties of cities.

**No stationary distribution for cities**

The equation (3) governs the evolution of urban population and analyzing it at large times gives indications about the stationary distribution of cities. In order to discuss analytical properties of this Eq. (3), we assume that Gaussian fluctuations are negligible compared to the Lévy noise and write $\eta_i \approx r$ (See Extended Data Figure 5). The corresponding Fokker-Planck equation (with Itô's convention) can be solved using the formalism of Fractional Order Derivatives and Fox functions [38, 39, 40, 41], leading to the general distribution at time t that can be expanded in powers of $S$ as (see SI for derivation and complete expressions of all terms):

$$P(S,t) = \sum_{k=1}^{\infty} C_k \frac{a(t)^{-\alpha\beta-\alpha(1-\beta)k}}{S^{1+\alpha\beta+\alpha(1-\beta)k}} \quad (4)$$

where $C_k$ is a complicated prefactor independent from time and $S$ and where $a(t) \propto \left[\frac{r/D^\alpha}{e^{r\alpha(1-\beta)t}-1}\right]^{1/\alpha(1-\beta)}$ decreases exponentially at large times. This expansion shows that the probability distribution of city sizes is dominated at large $S$ by the order $k = 1$ and converges towards a Pareto distribution with exponent $\alpha \neq 1$. The speed of convergence towards this power-law can be estimated with the ratio $\lambda(S, t)$ of the first and second terms of the expansion Eq. (4) and leads to:

$$\lambda(S,t) = \frac{D^\alpha}{r}\left(\acute{S}(t)/S\right)^{\alpha(1-\beta)} \quad (5)$$

where $\acute{S}(t)$ is the average city size. If $\lambda(S) \geq 1$, the $\alpha$-exponent regime is not valid in the right-tail with threshold $S$ at time $t$. Estimates of $\alpha$ and $\beta$ for the four datasets show that finite-time effects are very important in general and that a power-law regime is only reached for unrealistically large city sizes (see discussion in the SI for details). Hence, the range of city sizes for which we can observe a power-law distribution may not exist in practice and there is no reason in general to observe Zipf's law or any other stationary distribution. We also note that from Eq. 4 there is a scaling of the form $P(S,t) = \frac{1}{S}F\left(\frac{S}{S(t)}\right)$ with a scaling function $F$ that depends on the country. We confirmed this scaling form for France (the only country for which we had sufficient data) and details can be found in the SI (see Fig. S6).

In addition, if we perform a power-law fit of the expansion (Eq. 4), the upper-tail of the city size distributions may be mistaken for a Pareto-tail with a spurious exponent that changes with the definition of the upper-tail (Fig. S7 of Extended data). This might explain the various discrepancies observed in the broad literature about Zipf's law. As city sizes increase, the apparent exponent changes and can significantly deviate from 1 as we initially observed on the Extended Data Figure 1. Following our analysis, it should converge towards the value given by $\alpha$, as it is indeed observed for example for France ($\alpha = 1.4$) and the United States ($\alpha = 1.3$).

**Dynamics: splendor and decline of cities**

The validity of our model (Eq. (3)) can be further tested on the dynamics of systems of cities over large periods of time. This can be done by following the populations and ranks of the system's cities at different times with the help of "rank clocks" proposed in [5]. In this work, it was proven that the micro-dynamics of cities is very turbulent with many rises and falls of entire cities that cannot result from Gabaix's model (which is basically Gibrat with a non-zero minimum for city sizes). We show in Figure 2 the empirical rank clock for France (from 1876 to 2015) and for the results obtained with Gabaix's model and ours (for the other countries, see Extended Data Figure 8).

We see that in Gabaix's model (middle), city rank is on average stable and not turbulent: the rank trajectories are concentric and the rank of a city oscillates around its average position. In the real dynamics (left), cities can emerge or die. Very fast changes in rank order can occur, leading to a much more turbulent behavior. In our model (right), the large fluctuations of Lévy's noise are able to statistically reproduce such fast surges and swoops of cities. More quantitatively, we first compare the average shift per time $d = \frac{\sum_t \sum_{i=1}^N |r_i(t) - r_i(t-1)|}{NT}$ over $T$ years and for $N$ cities in the three cases (Table 3) and look at the statistical fluctuations of the rank (see Extended Data Figure 9): we note that Lévy fluctuations are much more able to reproduce the turbulent properties of the dynamics of cities through time. Indeed, the fast births or deaths of cities due for example to wars, discoveries of new resources, incentive settlement policies, etc. are statistically explained by broadly-distributed migrations and incompatible with a Gaussian noise. Second, we can also compare with the empirical data the predictions of the different models for the time needed to make the largest rank jump (see Extended data Figure 10 for France which typically predicts a duration of order 80 years to make a very large jump). We confirm here that Gabaix's model is unable to reproduce these very large fluctuations and that our equation agrees very well with the data.

**A new paradigm**

In this paper, we built a stochastic equation of growth for cities on microlevel considerations, empirically sound, that challenges the paradigm of Zipf's law and current models of urban growth. We show that microscopic details are irrelevant and the growth equation obtained is universal. A crucial point in this reasoning is that although we have on average some sort of detailed balance that would lead to a Gaussian multiplicative growth process, it is the existence of non-universal and broadly distributed fluctuations of the microscopic migration flows between cities that govern the statistics of city populations. We show here that city growth is described by a stochastic equation of a new type with two sources of noise and which predicts an asymptotic power-law regime. However, this stationary regime is not reached in general and finite-time effects cannot be discarded. Our model is also able to statistically reproduce the turbulent micro-dynamics of cities with fast births and deaths, in contrast to previous Gaussian-based models of growth [5].

In addition, our fundamental result exhibits an interesting connection between the behavior of complex systems and non-equilibrium statistical physics for which microscopic currents and the violation of detailed balance seem to be the rule rather than the exception [11]. At a practical level, this result also highlights the critical effect of not only inter-urban migration flows (an ingredient that is generally not considered in urban planning theories), but more importantly their large fluctuations, ultimately connected to the capacity of a city to attract a large number of new citizens. Our approach, that relies essentially on the population budget description and empirical results, provides a solid ground for future research on this central problem in urban science that is the temporal evolution of cities.

**Figure and Tables**

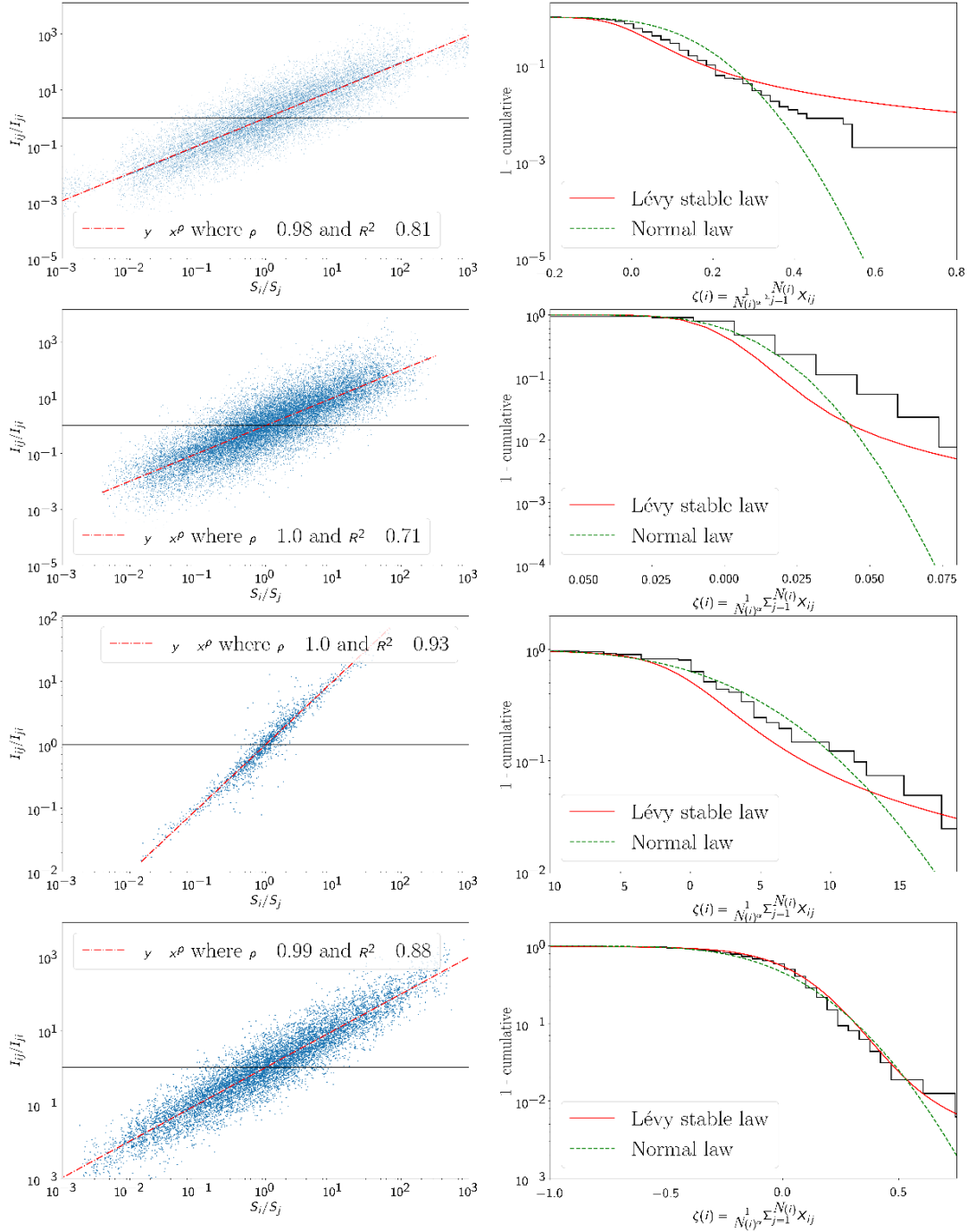

Figure 1. **Migration flow analysis.** From top to bottom: France, USA, UK, and Canada. Left: plot of the migration rate ratio versus the ratio of populations. The straight line is a power-law fit which gives an exponent equal to one. Right: Empirical right-cumulative distribution function of renormalized migrations flows $\zeta_i$ compared to Lévy (continuous red lines) and Normal distributions (green dashed lines). See Extended Data Figure 4 for the left-cumulative.

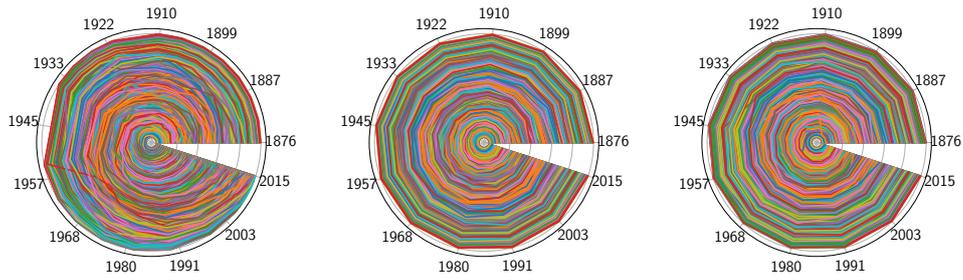

Figure 2. **Rank clocks for France.** We compare the real dynamics of the 500 largest French cities between 1876 and 2015 (left) to Gabaix's statistical prediction (middle) and to our statistical prediction (right). On the clocks, each line represents a city rank over time where the radius is given by the rank and the angle by time. In this representation, the largest city is at the center and the smallest at the edge of the disk.

**Tables**

| Dataset | MLE | Kolmogorov-Smirnov Test | Log-moments | Hill |
|---|---|---|---|---|
| France 2003-2008 | 1.43 ± 0.07 | 1.2 < $\alpha$ < 1.8 | 1.3 | 1.4 ± 0.3 |
| US 2013-2017 | 1.76 ± 0.07 | 1.7 < $\alpha$ < 1.8 | 1.6 | Inconclusive |
| UK 2012-2016 | 1.32 ± 0.26 | Inconclusive | 1.0 | 1.2 ± 0.8 |
| Canada 2012-2016 | 1.69 ± 0.12 | Inconclusive | 1.9 | 1.4 ± 0.6 |

Table 1. **Estimates of the parameter $\alpha$.** We used different methods of estimation: maximum likelihood estimation, Kolmogorov-Smirnov test, log-moments and Hill estimates (see for example [42]).

| Dataset | $\gamma$ | $\nu$ | $\beta = \nu + \gamma/\alpha$ | $\beta_{measured}$ |
|---|---|---|---|---|
| France 2003-2008 | 0.55 ± 0.06 | 0.4 ± 0.3 | 0.8 ± 0.4 | 0.75 ± 0.07 |
| US 2013-2017 | 0.36 ± 0.05 | 0.4 ± 0.4 | 0.6 ± 0.5 | 0.96 ± 0.05 |
| UK 2012-2016 | 0 | 0.7± 0.3 | 0.7± 0.3 | 0.51 ± 0.05 |
| Canada 2012-2016 | 0 | 0.5± 0.4 | 0.5± 0.4 | 0.78 ± 0.06 |

Table 2. **Estimates of parameters for the four datasets**. We observe a good agreement between the measured and predicted values of $\beta$ for France (see SI for details about these estimates). The large error bars (observed in particular for the US case) are due to the uncertainties on the value of $\nu$. British and Canadian datasets are small and hence fully-connected (implying $\gamma = 0$) and very noisy.

| Distance | Real | Lévy | Gabaix |
|---|---|---|---|
| France 1876-2015 | 6.0 | 6.1 | 8.0 |
| US 1790-1990 | 4.7 | 16 | 27 |
| UK 1861-1991 | 4.8 | 16 | 25 |

Table 3. **Average rank shift per unit time** $d$. Parameters for the Lévy and the Gaussian noise are fitted on the France 2003-2008, US 2013-2017 and UK 2012-2016 datasets respectively. The most complete dataset is the French one with total population for all cities at all times. In the US and UK datasets, only populations of the largest cities are recorded (Top 100 in the US and Top 40 in the UK). This can explain the large discrepancies observed while considering the distance $d$ in both countries. Extended Data Figure 10 investigates the distribution and fluctuations of the rank over time.

**Methods**

For the four countries we build the graph of migration flows between metropolitan areas. We have:
- The populations of metropolitan areas.
- The migration flows between metropolitan areas (described in more detail below).

**US migrations**
Data of migrations in the United States are taken from the 2013-2017 American Community Survey (ACS) [43]. Aggregated Metro Area to Metro Area Migration Flows and Counterflows are directly given between 389 metropolitan statistical areas (MSA) in the US. More precisely, the American Community Survey (ACS) asked respondents whether they lived in the same residence one year ago and for people who lived in a different residence, the location of their previous residence is collected.

**French interurban migrations**
Data of migrations in France are taken from the 2008 INSEE report for residential migrations at the town (commune) level for each individual household [44]. The main residence in 2008 is compared to the main residence in 2003. In order to work at the urban area level, we used the 1999 INSEE list of urban areas and aggregate residential migrations at the metropolitan level allowing us to analyze migration flows between the 500 largest urban areas in France.

**UK interurban migrations**
Data of migrations in the UK are taken from 2012-2016 ONS reports on internal migration between English and Welsh local authorities, giving the square matrix of moves each year [45]. In order to work at the urban area level, we used the list of local authorities by OECD functional urban areas and aggregate residential migrations at the metropolitan level allowing us to analyze migration flows between the 41 largest urban areas in England and Wales.

**Canadian interurban migrations**
Data of migrations in Canada are taken from 2012-2016 census reports on internal migration between Canadian metropolitan areas [46]. Flows between CMA are given city-to-city for each year between 2012 and 2016 for Canadian Top 160 cities.

**Data availability statement:** The datasets used in this study are freely available from public repositories: the American community survey for the US data [43] available at the address https://www.census.gov/data/tables/2017/demo/geographic-mobility/metro-to-metro-migration.html, accessed Feb. 21, 2020, the INSEE for the French data [44] available at the following link: https://www.insee.fr/fr/statistiques/2022291, accessed Feb. 21, 2020, the ONS data for the UK at https://www.ons.gov.uk/peoplepopulationandcommunity/populationandmigration/migrationwithintheuk/datasets/matricesofinternalmigrationmovesbetweenlocalauthoritiesandregionsincludingthecountriesofwalesscotlandandnorthernireland, accessed Jun 2, 2020 and the Canadian data at https://www150.statcan.gc.ca/t1/tbl1/en/tv.action?pid=1710008701&request_locale=en, accessed June 2, 2020.

**Acknowledgements:**
VV is supported by the Ecole Nationale des Ponts et Chaussees and by the Complex Systems Institute of Paris Île-de-France (ISC-PIF).

**Authors contributions:** VV and MB designed the study, VV acquired the data, VV and MB analyzed and interpreted the data and wrote the manuscript.

**Competing interests:** None.

**Supplementary Information is available for this paper.**